\documentclass[conference]{IEEEtran}
\IEEEoverridecommandlockouts
\usepackage{cite}
\usepackage{amsmath,amssymb,amsfonts}
\usepackage{graphicx}
\usepackage[dvipsnames]{xcolor}
\usepackage{xurl}
\usepackage{algorithm}
\usepackage{algpseudocode}
\usepackage{multirow}
\usepackage{makecell} 
\usepackage{fancyhdr}

\def\BibTeX{{\rm B\kern-.05em{\sc i\kern-.025em b}\kern-.08em
    T\kern-.1667em\lower.7ex\hbox{E}\kern-.125emX}}
\begin{document}

\title{Investigating Forecasting Models for Pandemic Infections Using Heterogeneous Data Sources: A 2-year Study with COVID-19\\}

\author{
        \IEEEauthorblockN{Zacharias Komodromos}
            \IEEEauthorblockA{
            \textit{KIOS Center of Excellence}\\
            \textit{University of Cyprus}\\
            Nicosia, Cyprus \\
            komodromos.zacharias@ucy.ac.cy\\
            ORCID: 0009-0002-8999-1201}
        \and
        \IEEEauthorblockN{Kleanthis Malialis}
            \IEEEauthorblockA{
            \textit{KIOS Center of Excellence}\\
            \textit{University of Cyprus}\\
            Nicosia, Cyprus \\
            malialis.kleanthis@ucy.ac.cy\\
            ORCID: 0000-0003-3432-7434}
        \and
        \IEEEauthorblockN{Panayiotis Kolios}
            \IEEEauthorblockA{
            \textit{KIOS Center of Excellence}\\
            \textit{Department of Computer Science}\\
            \textit{University of Cyprus}\\
            Nicosia, Cyprus \\
            kolios.panayiotis@ucy.ac.cy\\
            ORCID: 0000-0003-3981-993X}
            
        \thanks{This work was supported by the European Union’s Horizon 2020 research and innovation programme under grant agreement No 739551 (KIOS CoE - TEAMING) and from the Republic of Cyprus through the Deputy Ministry of Research, Innovation and Digital Policy. It was also supported by the CIPHIS (Cyprus Innovative Public Health ICT System) project of the NextGenerationEU programme under the Republic of Cyprus Recovery and Resilience Plan under grant agreement C1.1l2.}
}

\maketitle

\thispagestyle{plain}

\begin{abstract}
Emerging in December 2019, the COVID-19 pandemic caused widespread health, economic, and social disruptions. Rapid global transmission overwhelmed healthcare systems, resulting in high infection rates, hospitalisations, and fatalities. To minimise the spread, governments implemented several non-pharmaceutical interventions like lockdowns and travel restrictions. While effective in controlling transmission, these measures also posed significant economic and societal challenges. Although the WHO declared COVID-19 no longer a global health emergency in May 2023, its impact persists, shaping public health strategies. The vast amount of data collected during the pandemic offers valuable insights into disease dynamics, transmission, and intervention effectiveness. Leveraging these insights can improve forecasting models, enhancing preparedness and response to future outbreaks while mitigating their social and economic impact. This paper presents a large-scale case study on COVID-19 forecasting in Cyprus, utilising a two-year dataset that integrates epidemiological data, vaccination records, policy measures, and weather conditions. We analyse infection trends, assess forecasting performance, and examine the influence of external factors on disease dynamics. The insights gained contribute to improved pandemic preparedness and response strategies.

\indent \textit{Clinical relevance}—This study relies on anonymised, aggregated epidemiological data, including total infections, hospitalisations, ICU admissions, and deaths, rather than individual patient tracking. Our findings could potentially contribute to healthcare by improving forecasting models that help hospitals anticipate surges, allocate resources more efficiently, and prevent system overload. By analysing the effects of policy interventions and external factors such as weather conditions, this research may provide valuable insights for refining public health strategies. Beyond COVID-19, our approach could be used to enhance infectious disease forecasting, supporting proactive decision-making in future outbreaks.

\end{abstract}


\section{Introduction}
\vspace{-2pt}
A pandemic is defined by the World Health Organisation (WHO) as an exceptional occurrence that poses a public health risk to other nations due to the international spread of a disease and may necessitate a coordinated global response \cite{who2019}. This definition highlights the severity and global impact of pandemics, which have historically posed significant challenges to public health and economic stability. Evidence suggests that the frequency of pandemics has increased in recent decades \cite{jones2008global}, necessitating more robust monitoring and forecasting mechanisms.

COVID-19 became one of the most disruptive health crises in modern history \cite{coronaviridae2020species}. The virus spread rapidly across the globe, overwhelming healthcare systems and leading to an unprecedented number of infections, hospitalisations, and fatalities. Beyond health, it caused severe economic and social disruptions. Governments worldwide have implemented non-pharmaceutical interventions, such as lockdowns, and social distancing, to contain the spread. These measures, while effective in controlling transmission, have also created major challenges for economic stability and public well-being.

On May 5, 2023, the WHO declared that COVID-19 was no longer a global health emergency, marking a significant shift in the pandemic’s trajectory \cite{who2023}. However, COVID-19’s impact remains substantial, and its legacy continues to influence global public health strategies. Since its onset, an unprecedented amount of data has been collected from diverse sources. The wealth of information gathered during the COVID-19 pandemic presents an opportunity to enhance our understanding of infectious disease forecasting. These data provide valuable insights into disease dynamics, transmission patterns, and intervention effectiveness. Lessons learned from COVID-19 can be leveraged to develop more accurate predictive models, which can be critical in managing future outbreaks. By refining forecasting methods and improving preparedness strategies, researchers and policymakers can ensure a more effective response to emerging health threats, ultimately reducing the social and economic toll of future pandemics.

The key contributions of this work are as follows.
\begin{enumerate}
    \item We conduct a large-scale case study on Cyprus, a European country with a population of nearly 1 million. Our dataset spans two years and integrates diverse sources, including epidemiological data (infections, hospitalisations, intensive care unit (ICU) data, deaths), vaccination records, policy interventions, and weather conditions.

    \item We focus on forecasting COVID-19 infections (i.e., positive cases), presenting key findings and insights into the dynamics of the pandemic, along with an evaluation of forecasting performance. We assess the impact of various factors, such as policy measures and weather conditions, on infection trends thus offering a deeper understanding of their role in pandemic evolution.

    \item Our findings provide valuable insights for patients, healthcare professionals, and policymakers, contributing to better preparedness and response strategies for future pandemics. While centered on Cyprus, our approach and conclusions may be relevant to other countries as well.
\end{enumerate}

The rest of the paper is structured as follows. Section~\ref{sec:related} describes related work. Section~\ref{sec:methods} describes the datasets used in this study, and presents the methods used, including, feature extraction methods, compared learning and statistical methods, and evaluation method. Section~\ref{sec:results} discusses the findings and results of our study. Section~\ref{sec:conclusion} concludes this work.

\section{Related Work}\label{sec:related}

The task of forecasting pandemic trends has been extensively studied, with various approaches applied to predict COVID-19 outcomes, mainly new cases, hospitalisations, and deaths \cite{ayoobi2021time}. These methods can be broadly classified into two categories, compartmental models and data-driven methods.

\par \textbf{Compartmental models} are fundamental in epidemiology for studying infectious diseases. These models operate by dividing the total population into distinct compartments based on disease status. Mathematically, they rely on systems of ordinary differential equations (ODEs), where each equation describes the rate of change in the number of individuals in each compartment over time.

Two well-adopted models include the Susceptible-Infected-Recovered (SIR) \cite{nesteruk2020simulations_sir} and the Susceptible-Exposed-Infected-Recovered (SEIR) models \cite{pandey2020seir}. Beyond classical compartmental models, more complex ones have been developed to better capture COVID-19 dynamics. For instance, the `SIDAREVH' model \cite{karapitta2024time} \cite{karapitta2024pandemic} introduces additional compartments that distinguish between vaccinated and unvaccinated individuals at different disease stages, specifically separating infected detected cases as well as hospitalised cases based on vaccination status.

\par \textbf{Data-driven methods}. Data-driven approaches have been extensively employed for forecasting COVID-19 trends, leveraging both traditional statistical techniques and more advanced machine learning and deep learning models.

Common statistical models, such as ARMA and ARIMA \cite{alzahrani2020forecasting_arma_arima}\cite{benvenuto2020application_arima} have been widely used due to their simplicity and effectiveness in time series forecasting. These models are well-suited for short-term predictions and stationary datasets but often face difficulties when applied to non-stationary\cite{ryan2024_non_stationarity} pandemic data. Changes in transmission dynamics, public health policies, tourist arrivals, and social behaviours introduce complexities that can limit the applicability of purely statistical approaches in highly dynamic environments.

Traditional machine learning techniques, such as Simple Linear Regression \cite{ogundokun2020predictive_linear_regression}, and tree-based algorithms, such as XGBoost \cite{fang2022application_xgboost}, have been widely applied in epidemiological forecasting. These models offer interpretability and flexibility in handling structured datasets, as they can incorporate engineered features that capture pandemic dynamics.

Deep learning models, e.g., Long Short-Term Memory (LSTM) and Gated Recurrent Unit (GRU) networks \cite{chimmula2020time_lstm, shahid2020predictions_comparative1}, and Variational AutoEncoder (VAE) \cite{zeroual2020deep_comparative2} have demonstrated strong capabilities in capturing non-linear dependencies and long-term temporal patterns. These models excel at processing time series data while reducing the need for extensive manual feature engineering. However, deep learning methods often require large volumes of data, making them less practical in the early stages of an outbreak or in regions with limited historical data. They are also generally considered less interpretable.

Lastly, the paradigm of online incremental learning \cite{malialis2020online} has been used to capture changes in the data distribution (i.e., concept drift) in the spread of COVID-19, and adapt models in real time \cite{stylianides2023study, tetteroo2022automated_onlineextra1}.

\section{Data and Methods}\label{sec:methods}

\subsection{Datasets}\label{sec:datasets}
The primary dataset consists of daily SARS-CoV-2 positive cases in Cyprus, covering the period from 1st October 2020 to 31st December 2022, along with key epidemiological indicators such as hospitalisations, deaths, ICU admissions, and intubations. These data were publicly available data collected by the MoH of the Republic of Cyprus and published during the evolution of the pandemic in weekly reports by the Press and Information Office.

\begin{figure}[t]
    \centering
    \includegraphics[width=0.49\textwidth]{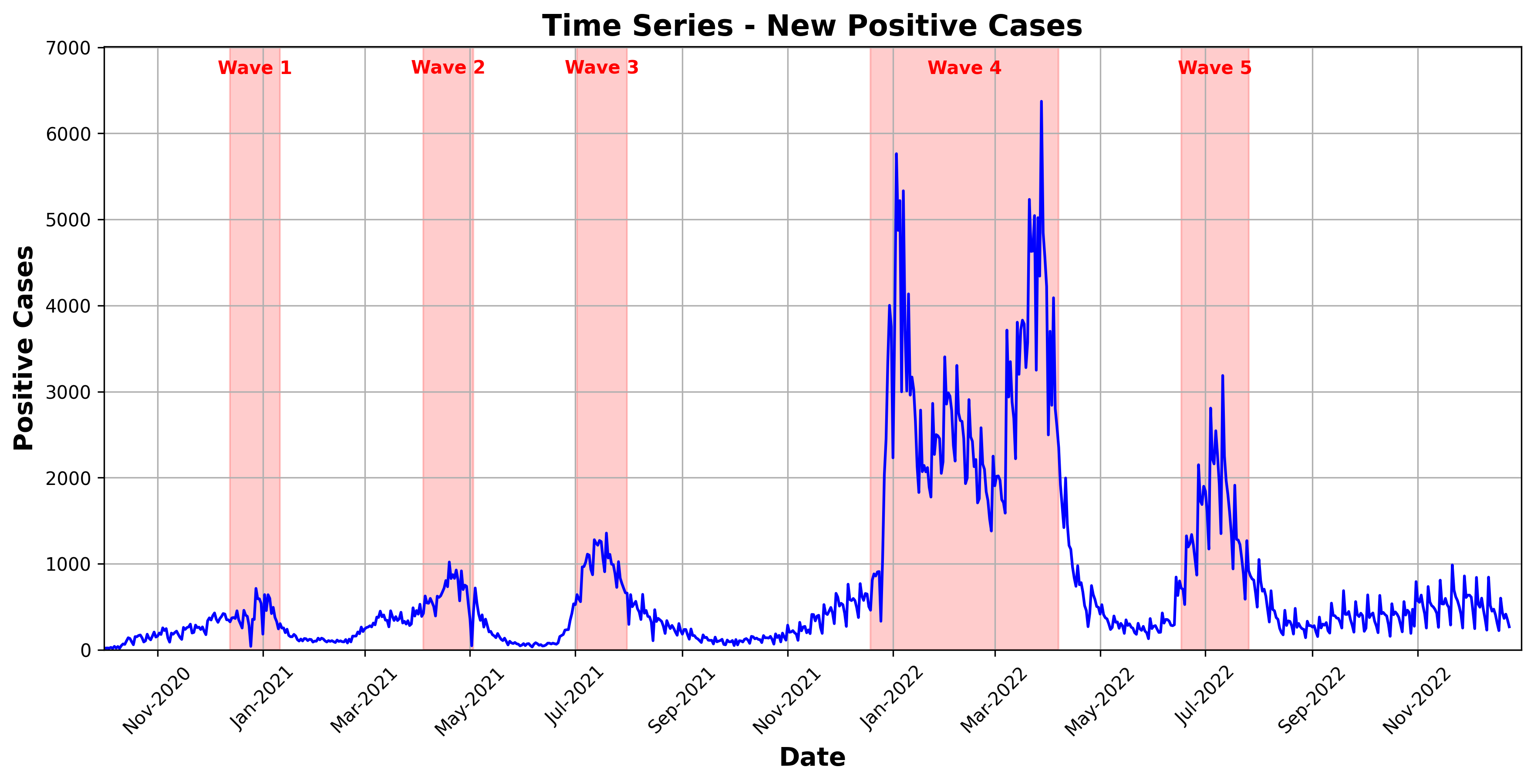}
    \vspace{-20pt}
    \caption{Daily COVID-19 cases in Cyprus (01/10/2020 to 31/12/2022), highlighting five distinct waves.}
    \label{fig:waves_cases}
    \vspace{-17pt}
\end{figure}

\textbf{Infection data}.
During the study period, Cyprus experienced five distinct pandemic waves, which are defined as follows and depicted in Figure~\ref{fig:waves_cases}.
\begin{itemize}
    \item \textbf{Wave 1:} 13th December 2020 – 11th January 2021
    \item \textbf{Wave 2:} 4th April 2021 – 3rd May 2021
    \item \textbf{Wave 3:} 2nd July 2021 – 31st July 2021
    \item \textbf{Wave 4:} 19th December 2021 – 7th April 2022
    \item \textbf{Wave 5:} 17th June 2022 – 26th July 2022
\end{itemize}

A diverse set of information from multiple datasets has also been considered, as described below.

\begin{figure*}[t]
    \centering
    \includegraphics[width=\textwidth]{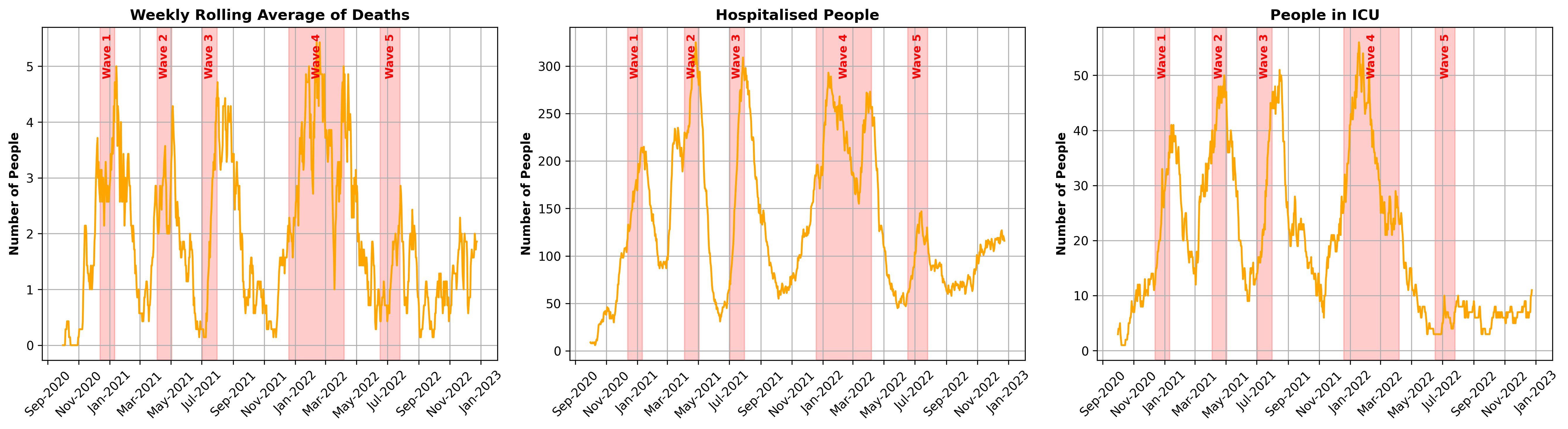}
    \vspace{-20pt}
    \caption{Time series of (1) weekly rolling average of deaths, (2) total of hospitalised people due to COVID-19 per day, and (3) total people in ICU due to COVID-19 per day.}
    \label{fig:snow_3subplots}
    \vspace{-11pt}
\end{figure*}

\textbf{Epidemiological data}.
Daily case reports, hospital admissions, ICU occupancy, intubations, and deaths were collected to track the evolution of COVID-19 in Cyprus. These indicators provide a comprehensive view of the disease severity and the burden on the healthcare system. Figure~\ref{fig:snow_3subplots} presents the weekly rolling average of deaths alongside the number of hospitalised patients and those in intensive care units (ICU). As expected, deaths are strongly correlated with hospitalisations and ICU admissions, with Pearson’s correlation coefficients of 0.74 and 0.71, respectively, when calculated between the weekly moving average of deaths and each field, indicating a strong positive correlation. However, deaths do not exhibit a high correlation with daily positive cases (Figure~\ref{fig:waves_cases}). The Pearson’s correlation coefficient between daily deaths and daily infections is 0.31, suggesting a weak positive correlation, while the correlation between weekly rolling averages of deaths and infections is moderate at 0.46. Moreover, fluctuations in case volumes across different waves are not proportional to trends in deaths and hospitalisations, as the latter reach similarly high peaks during their respective surges.

\begin{figure}[t]
    \centering
    \includegraphics[width=0.45\textwidth]{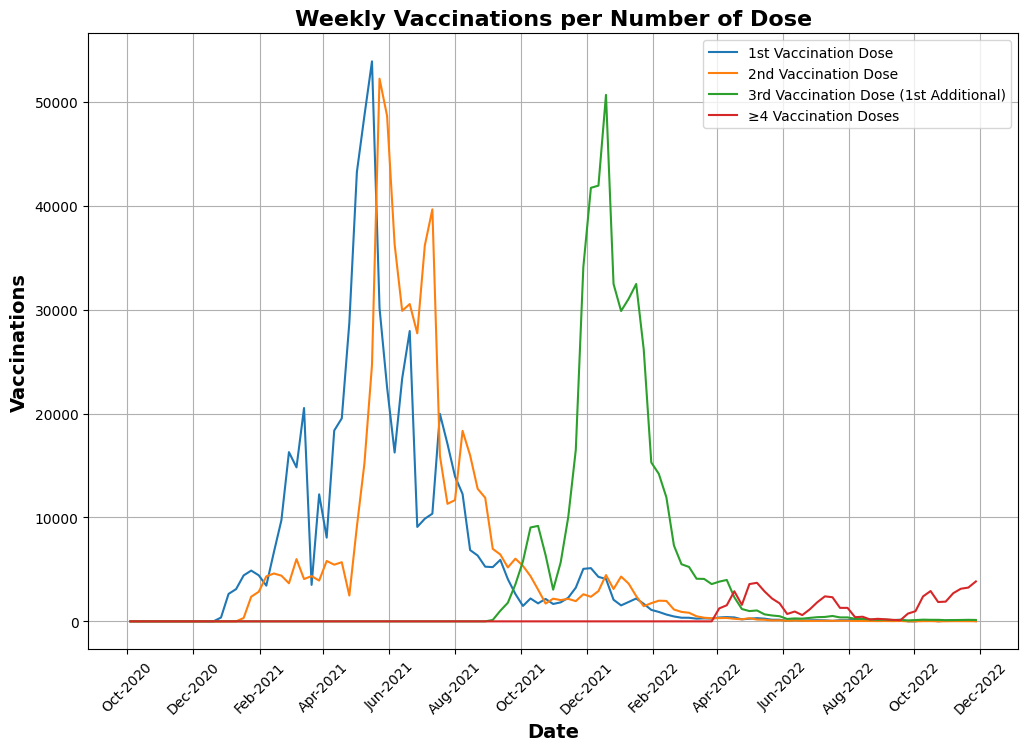}
    \vspace{-10pt}
    \caption{Time series of weekly vaccinations per number of dose.}
    \label{fig:vaccinations}
    \vspace{-15pt}
\end{figure}

\textbf{Vaccination data}.
Vaccination records were analysed to assess their role in short-term forecasting. These data are available through the National Open Data Portal of Cyprus \cite{vaccinations}. Figure~\ref{fig:vaccinations} illustrates the weekly number of vaccinations per dose category. The volume of first, second, and third doses follows a similar trend within their respective time periods, whereas the uptake of the fourth and subsequent doses is significantly lower.

\begin{figure*}[!t]
    \centering
    \includegraphics[width=\textwidth]{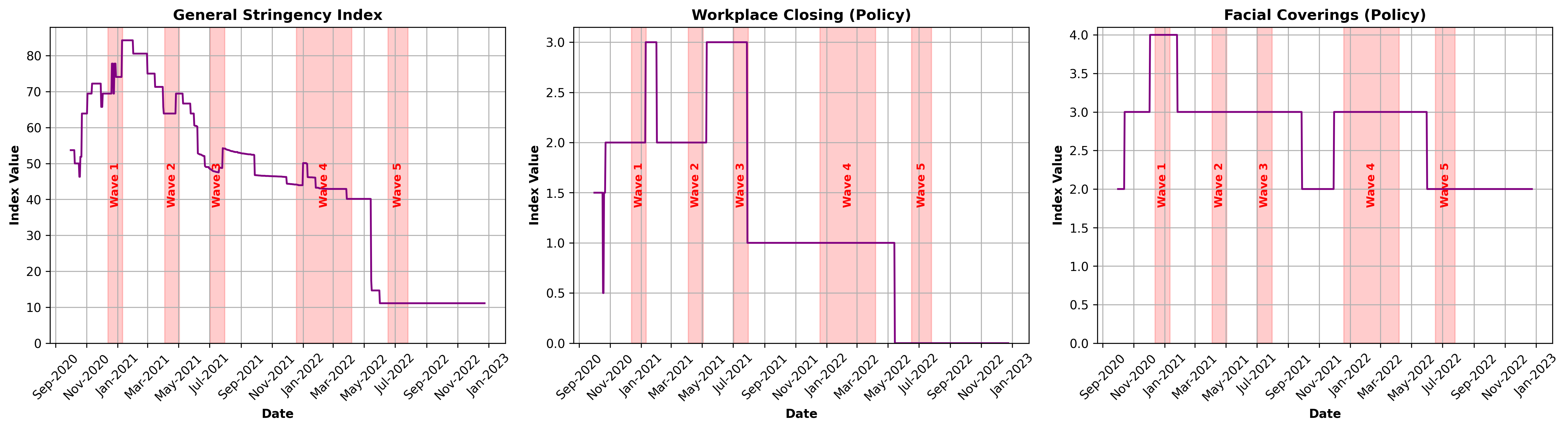}
    \vspace{-20pt}
    \caption{Time series of daily policy indices. In order: General Stringency Index, Workplace Closing, and Facial Coverings.}
    \label{fig:policies_3subplots}
    \vspace{-15pt} 
\end{figure*}

\textbf{Policy data}. The Oxford COVID-19 Government Response Tracker (OxCGRT) \cite{policies} was used to capture policy indices reflecting government interventions, such as lockdowns and travel restrictions. Figure~\ref{fig:policies_3subplots} depicts the general stringency index along with workplace closing, and facial coverings policies. After May 2022, the indices do not change further, which may indicate either a discontinuation in data collection or sustained stability in policy measures beyond that point. The dataset includes a total of 15 other unique policy indices which are not displayed due to space restrictions.

\begin{figure}[t]
    \centering
    \includegraphics[width=0.48\textwidth]{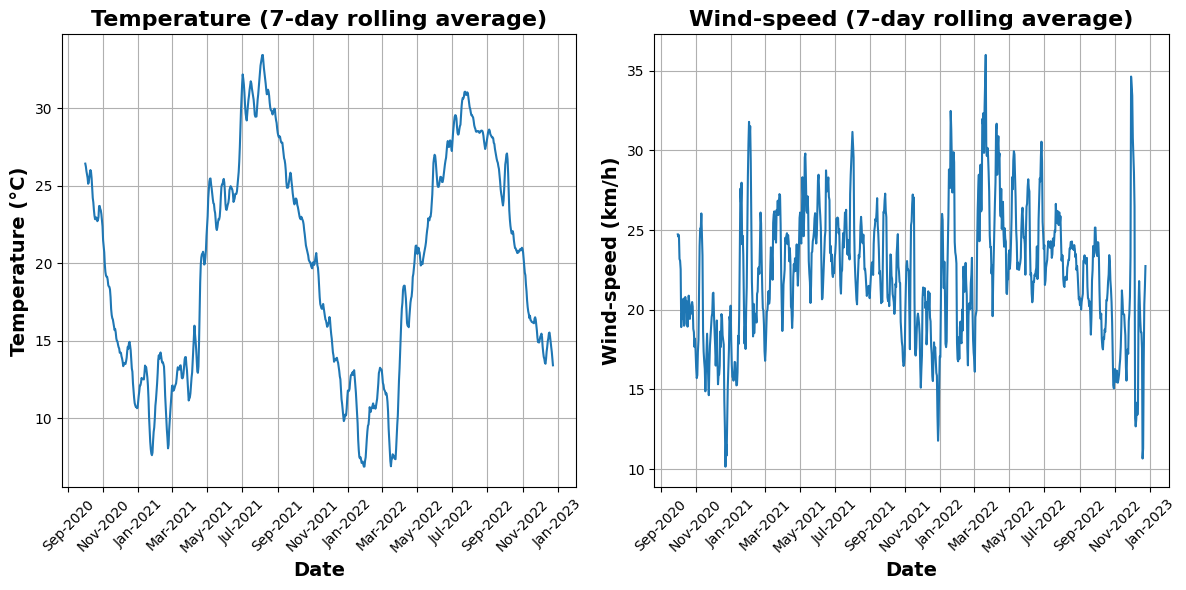}
    \vspace{-10pt}
    \caption{Weekly rolling averages of average daily temperature and average daily wind-speed.}
    \label{fig:weather}
    \vspace{-15pt} 
\end{figure}

\textbf{Weather data}. Environmental factors such as temperature, humidity, and wind speed were included to examine potential correlations between climatic conditions and infection rates. The weather data were obtained from Visual Crossing \cite{weather_data}. Figure~\ref{fig:weather} illustrates the weekly rolling average of the daily mean temperature, and wind-speed.

\begin{table*}[t]
\centering
\caption{Feature Categories and their Corresponding Extracted Features. Selected Features are coloured.}
\vspace{-5pt} 
\label{tab:feature_categories}
\renewcommand{\arraystretch}{1.3}
\begin{tabular}{|m{3cm}|p{14cm}|} 
\hline
\textbf{Feature Category} & \textbf{Features} \\ \hline
\textbf{Infections (Raw)} & 
\textcolor{blue}{Lagged 7-day windows of positive cases}.\\ \hline

\textbf{Infections (Features)} & 
\textcolor{blue}{Trend indicators (upward/downward indicators, positive cases ratio comparing current week with previous week), 7-day moving average, days of the week}, positive cases by age group (0-1, 2-14, \textcolor{blue}{15-24, 25-44}, 45-64, 65+).\\ \hline

\textbf{Epidemiological} & 
\textcolor{OliveGreen}{New hospitalisations, total hospitalised patients, new ICU admissions, total ICU patients}, hospitalised patients categorised by days since admission (1-3 days, 4-7 days, 1-2 weeks, 2+ weeks), ICU patients categorised by days since admission (1-3 days, 4-7 days, 1-2 weeks, 2+ weeks), new intubations, total intubated patients, intubated patients categorised by days since intubation (1-3 days, 4-7 days, 1-2 weeks, 2+ weeks), lagged 7-day windows (hospitalisations, deaths). \\ \hline

\textbf{Vaccinations} & 
\textcolor{OliveGreen}{Average daily vaccinations computed from weekly totals}, partition of the average daily vaccinations of the week, into first, second, third, and fourth+ doses, as well as their cumulative totals.
 \\ \hline

\textbf{Policies} & 
\textcolor{OliveGreen}{Overall stringency index}, government restrictions including school closures, workplace closures, public event cancellations, gathering restrictions, stay-at-home requirements, internal and international travel restrictions, mask mandates, vaccination policies. \\ \hline

\textbf{Weather} & 
\textcolor{OliveGreen}{7-day moving averages of temperature, humidity, wind speed} and precipitation, daily average values of temperature, humidity, wind speed and precipitation. \\ \hline
\end{tabular}
\vspace{-11pt} 
\end{table*}

\vspace{-5pt} 

\subsection{Feature extraction}
Table~\ref{tab:feature_categories} presents an overview of the raw features (Sec.~\ref{sec:datasets}), along with features extracted using the techniques described below. 
The colour coding refers to the feature selection process which is described in Section~\ref{sec:feature_selection}.

We retain infection counts and their corresponding lagged values, restricted to a one-week history. By focusing on recent data, we reduce noise from older observations and retain the most relevant signals for near-term forecasts, given the short-term autocorrelation in COVID-19 infections. To capture demographic effects, infections per age group have been introduced to account for age-specific influences on transmission trends. Additionally, weekly moving averages were computed to smooth fluctuations in features such as positive cases and weather variables. Furthermore, upward and downward trend indicators were introduced, along with a positive case ratio metric that compares the current week's infections to those of the previous week, offering insights into evolving pandemic dynamics. Finally, the days of the week were included as features, as testing demonstrated their usefulness in improving predictive performance.

\textbf{Data pre-processing}. 
A log transformation is applied to infection-related data, both for features and targets, to smooth fluctuations while retaining outliers. It effectively handles extreme values and exponential growth, while minimising variance by compressing the range and addressing right-skewness. One-hot encoding is applied to handle categorical variables.
Further data pre-processing methods specific to models
are described in the next section.

\subsection{Forecasting models}
Our study examines an advanced learning model (XGBoost) and a classical statistical model (ARIMAX).

\textbf{XGBoost}. Extreme Gradient Boosting (XGBoost) is a scalable and efficient tree-based ensemble learning method that builds upon gradient boosting by incorporating regularisation and optimisation techniques to enhance performance. It is known for its speed, predictive accuracy, and is particularly effective in time series forecasting \cite{chen2016xgboost}.

\textbf{ARIMAX}. The Autoregressive Integrated Moving Average with Exogenous Variables (ARIMAX) model extends ARIMA by incorporating external predictors to improve forecasting accuracy \cite{box2013box}. This makes ARIMAX particularly useful in epidemiological forecasting, where external factors such as weather conditions, can impact disease spread.

ARIMAX assumes stationarity, meaning that the key statistics like mean, variance, and autocovariance remain constant over time. However, real-world data often exhibit non-stationarity, typically due to trends or seasonality. To address this, we applied a two-step differencing procedure to each non-stationary exogenous variable. Let $y_t$ denote a non-stationary exogenous feature at time $t$. We first applied seasonal differencing with a lag of 7 days to remove weekly seasonality, given by $z_t = y_t - y_{t-7}$. This was followed by first-order differencing to remove any remaining linear trend, given by $\Delta z_t = z_t - z_{t-1}$, which simplifies to the combined expression $(y_t - y_{t-1}) - (y_{t-7} - y_{t-8})$.

\subsection{Feature selection}\label{sec:feature_selection}
To reduce model complexity and improve predictive performance, we have performed feature selection strategies.

First, we consider the outcomes of two selection methods, Recursive Feature Elimination (RFE) \cite{guyon2002gene_rfe} and XGBoost's native Top Gain Ranking \cite{chen2016xgboost}, where a trial-and-error approach is employed to assess the impact of different feature subsets.

Second, a manual selection process has been followed. Regarding the infection data, we select specific features by removing those with a very high correlation, such as ICU cases and intubations which have a correlation of 0.97. Regarding the vaccination data, we use the average daily vaccinations computed from weekly totals, without considering the specific dose number. As shown in Figure~\ref{fig:vaccinations}, each dose follows a distinct timeline, leading to inconsistencies in the model, particularly because the first and second doses mainly fall within the training set but not the test set. Regarding the policy data, in Figure~\ref{fig:policies_3subplots} we observe that individual policy indices exhibit prolonged periods of steadiness with minimal variability. This resulted in low feature importance scores during selection tests. Therefore, to reduce noise, we consider only the general stringency index as a representative policy feature. Regarding the weather variables considered include the smoothed 7-day moving averages of temperature, humidity, and wind speed. Precipitation is excluded as it provides limited information due to the high frequency of dry days.

In conclusion, the most predictive features per category are shown in Table~\ref{tab:feature_categories} with a coloured font. Specifically, \textcolor{blue}{blue} indicates the features directly related to the task at hand, i.e, infections (13), along with the days of the week, while \textcolor{OliveGreen}{green} represents the variables (9) whose role and importance is examined further in Section~\ref{sec:results}.

\subsection{Evaluation method and performance metrics}

\begin{figure*}[t]
    \centering
    \includegraphics[width=\textwidth]{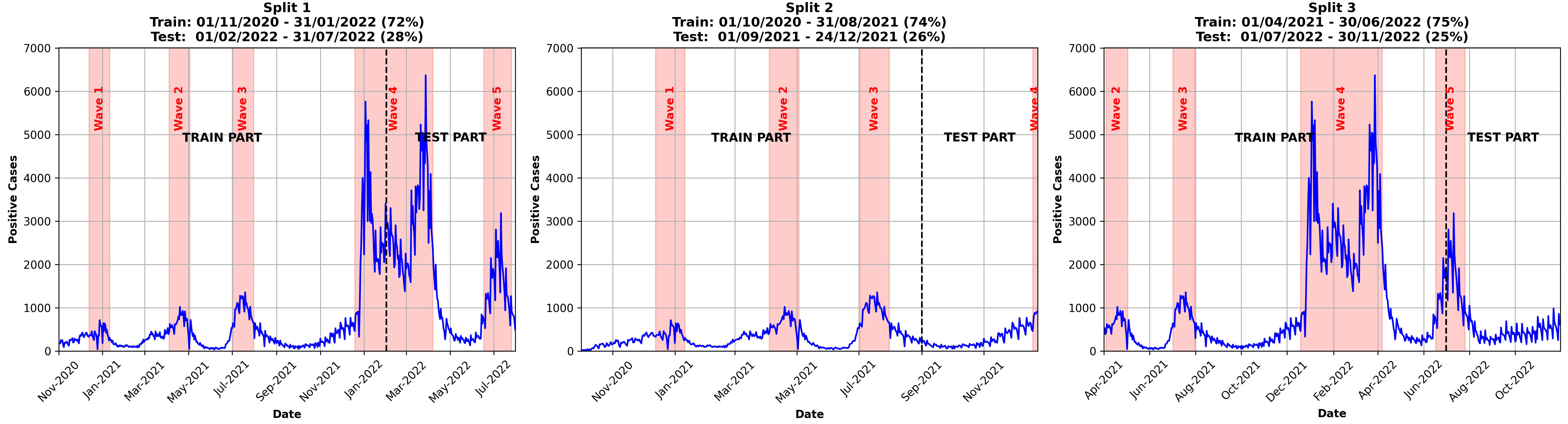}
    \vspace{-20pt}
    \caption{Training and test sets splits used for applications.}
    \label{fig:splits}
    \vspace{-15pt} 
\end{figure*}

\textbf{Training splits.} Figure~\ref{fig:splits} illustrates the selected data splits for our study period. The first split contains half of the largest wave along with the final wave, allowing us to evaluate the model’s performance in both waves while leveraging prior information on the largest wave. The second split consists of the three smaller waves in the training set, focusing on assessing the model’s effectiveness during low-volume periods and capturing only a few days of the fourth wave. The final split incorporates data from later stages when infection volumes begin to rise, enabling an evaluation of model performance during the last phase of the study, which includes the majority of the final wave.

\textbf{Overfitting control.}
To avoid overfitting, the time series cross-validation strategy using rolling splits was used, to guarantee that the model tuning respected temporal order and avoided lookahead bias. Hyperparameter optimisation focused on balancing accuracy and generalisation, for example by using regularisation in XGBoost to prevent unnecessary complexity. As discussed previously, feature selection techniques, including the removal of highly correlated features, were applied to decrease redundancy and improve model robustness.

To evaluate and compare the proposed methods, we utilise widely accepted regression metrics.

\textbf{MAPE.} The Mean Absolute Percentage Error (MAPE) is selected as the primary evaluation metric, as it accounts for variations in scale across different periods, such as high waves and low case periods. MAPE provides an intuitive interpretation of forecasting accuracy by expressing errors as percentages of actual values. It is computed as follows:
\vspace{-5pt}

\begin{equation}
\text{MAPE}_d = \frac{100\%}{n} \sum_{i=1}^{n} \left| \frac{y_{i,d} - \hat{y}_{i,d}}{y_{i,d}} \right|,
\label{eq:mape}
\end{equation}
\noindent where \(y_{i,d}\) denotes the true value for the \(i\)th observation at the \(d\)\textsuperscript{th} day ahead forecast, \(\hat{y}_{i,d}\) is the corresponding predicted value, and \(n\) is the total number of predictions.

\textbf{MAE.} The Mean Absolute Error (MAE) is also reported to provide an absolute measure of error. Unlike MAPE, MAE does not scale with the magnitude of values, making it more sensitive in periods of high case fluctuations. It is defined as:
\vspace{-5pt}

\begin{equation}
\text{MAE}_d = \frac{1}{n} \sum_{i=1}^{n} \left| y_{i,d} - \hat{y}_{i,d} \right|.
\label{eq:mae}
\end{equation}

Since we generate multi-step forecasts \(H\) days into the future, we also compute metrics that summarise model performance across the entire horizon. Here, \(H\) denotes the forecasting horizon, i.e., the number of days ahead being predicted. The averaged error metrics across this horizon are defined below:
\vspace{-5pt}

\begin{equation}
\text{MAPE}_{\text{H-days avg}} = \frac{1}{H} \sum_{d=1}^{H} \text{MAPE}_d ,
\label{eq:mape_Hdays_avg}
\end{equation}
where \(\text{MAPE}_d\) is the MAPE of the \(d\)\textsuperscript{th} day ahead, as computed in equation~\eqref{eq:mape}. 

The corresponding \(\text{MAE}_\text{H-days avg}\) is defined analogously by averaging the daily MAEs over the forecasting horizon.

For aggregating the results, we compare the sum of predicted values against the sum of actual values. The metrics are computed as:
\vspace{-5pt}

\begin{equation}
\text{MAPE}_{\text{H-days agg.}} = \frac{100\%}{n} \sum_{i=1}^{n} \left| \frac{\sum_{j=1}^{H} y_{i,j} - \sum_{j=1}^{H} \hat{y}_{i,j}}{\sum_{j=1}^{H} y_{i,j}} \right|,
\label{eq:mape_Hdays_aggr}
\end{equation}
\vspace{-5pt}

\begin{equation}
\text{MAE}_{\text{H-days agg.}} = \frac{1}{n} \sum_{i=1}^{n} \left| \sum_{j=1}^{H} y_{i,j} - \sum_{j=1}^{H} \hat{y}_{i,j} \right|.
\label{eq:mae_Hdays_aggr}
\end{equation}

For all XGBoost experiments, each model is trained and evaluated over 10 repetitions. The reported results include both the mean and standard deviation of these metrics, covering the entire time period as well as separate evaluations during wave and normal periods.

\section{Results and Discussion}\label{sec:results}

\subsection{Role of feature category on XGBoost}\label{sec:ablation}
This section examines the role and importance of each feature category described in Table~\ref{tab:feature_categories} on XGBoost's performance. In subsequent tables we refer to ``Features'', to all selected features from Table~\ref{tab:feature_categories} which are coloured in blue and green, based on the selection process described in Section~\ref{sec:feature_selection}.

\subsubsection{Role of infections (features)}
Table~\ref{tab:excluding_infections_features_days} shows XGBoost's performance with the selected features (`Features'), and its performance when excluding the infection data with features (`w/o Infections (Feat.)'). The results show a significant deterioration in model performance across all three splits when the second category of Table~\ref{tab:feature_categories} is removed. This is attributed to the fact that the features derived from the raw infection data, along with the days of the week, provide valuable information to XGBoost and increase its predictive power.

\begin{table}[t]
\centering
\caption{XGBoost's Perf. (MAPE) on Test Set using Selected Features Vs Excluding Infections (Feat.)}
\vspace{-5pt} 
\label{tab:excluding_infections_features_days}
\renewcommand{\arraystretch}{1.2}
\setlength{\tabcolsep}{10pt}

\resizebox{\columnwidth}{!}{
\begin{tabular}{|c|c|c|c|}
\hline
\textbf{Split} & \textbf{MAPE (\%)} & 
\makecell{\textbf{Features}} & 
\makecell{\textbf{w/o Infections (Feat.)} } \\ 
\hline
\multirow{2}{*}{1st} & 7-days avg  & \textbf{24.9} (2.1)  &  35.4 (3.2) \\ 
& 7-days agg.  & \textbf{20.7} (0.9)  & 27.3 (0.9) \\ 
\hline
\multirow{2}{*}{2nd} & 7-days avg  & \textbf{26.1} (0.7)  & 35.7 (1.2) \\ 
& 7-days agg.  & \textbf{14.5} (0.4)  & 24.2 (0.6) \\ 
\hline
\multirow{2}{*}{3rd} & 7-days avg  & \textbf{25.0} (1.3)  & 36.0 (2.1) \\ 
& 7-days agg.  & \textbf{17.8} (0.6)  & 23.8 (1.5) \\ 
\hline
\end{tabular}}
\vspace{-13pt} 
\end{table}

\subsubsection{Role of epidemiological data}
Table~\ref{tab:excluding_other_epidemiological_results} shows XGBoost's performance with the selected features (`Features'), and its performance when excluding the epidemiological data (`w/o Epidemiological'). Based on the observations, we conclude that incorporating additional epidemiological data improves XGBoost's predictive performance, and these data should therefore be retained.

\begin{table}[t]
\centering
\caption{XGBoost's Perf. (MAPE) on Test Set using Selected Features Vs Excluding Epidemiological Data}
\vspace{-5pt} 
\label{tab:excluding_other_epidemiological_results}
\renewcommand{\arraystretch}{1.2}
\setlength{\tabcolsep}{10pt}

\resizebox{\columnwidth}{!}{
\begin{tabular}{|c|c|c|c|}
\hline
\textbf{Split} & \textbf{MAPE (\%)} & 
\makecell{\textbf{Features}} & 
\makecell{\textbf{w/o Epidemiological} } \\ 
\hline
\multirow{2}{*}{1st} & 7-days avg  & \textbf{24.9} (2.1)  & 25.1 (1.8)  \\ 
& 7-days agg.  & \textbf{20.7} (0.9)  & 21.6 (0.9)  \\ 
\hline
\multirow{2}{*}{2nd} & 7-days avg  & \textbf{26.1} (0.7)  & \textbf{26.1} (0.7)  \\ 
& 7-days agg.  & \textbf{14.5} (0.4)  & 14.8 (0.3)  \\ 
\hline
\multirow{2}{*}{3rd} & 7-days avg  & \textbf{25.0} (1.3)  & 29.9 (2.6)  \\ 
& 7-days agg.  & \textbf{17.8} (0.6)  & 22.2 (1.5)  \\ 
\hline
\end{tabular}}
\vspace{-10pt} 
\end{table}

\subsubsection{Role of vaccination data}
Table~\ref{tab:excluding_total_vaccinations_results} shows XGBoost's performance with the selected features (`Features'), and its performance when excluding the average daily vaccinations computed from weekly totals (`w/o Vaccinations'). Vaccination data were found not to have predictive power, as they do not reduce the MAPE of our XGBoost model. We attribute this to the following reasons. First, vaccines typically take weeks to generate an immune response \cite{cdc2024}, and therefore the impact of the vaccines cannot be captured for the upcoming week, where our predictions lie.
Second, the COVID-19 vaccines have been proven highly effective in reducing severe illness, hospitalisations, and deaths, but their ability to prevent infections has been more limited, especially with the emergence of new variants \cite{zeng2022effectiveness, who_vaccinations_2021}. Third, the vaccination data considered in this study were in an aggregated format.

\begin{table}[t]
\centering
\caption{XGBoost's Perf. (MAPE) on Test Set using Selected Features Vs Excluding Vaccination data}
\vspace{-5pt} 
\label{tab:excluding_total_vaccinations_results}

\renewcommand{\arraystretch}{1.2}
\setlength{\tabcolsep}{11.4pt}

\resizebox{\columnwidth}{!}{
\begin{tabular}{|c|c|c|c|}
\hline
\textbf{Split} & \textbf{MAPE (\%)} & 
\makecell{\textbf{Features}} & 
\makecell{\textbf{w/o Vaccinations}} \\ 
\hline
\multirow{2}{*}{1st} & 7-days avg  & 24.9 (2.1)  & \textbf{23.0} (1.2)  \\ 
& 7-days agg.  & 20.7 (0.9)  & \textbf{17.7} (0.6)  \\ 
\hline
\multirow{2}{*}{2nd} & 7-days avg  & 26.1 (0.7)  & \textbf{25.9} (0.7)  \\ 
& 7-days agg.  & \textbf{14.5} (0.4)  & 14.6 (0.2)  \\ 
\hline
\multirow{2}{*}{3rd} & 7-days avg  & 25.0 (1.3)  & \textbf{22.8} (0.9)  \\ 
& 7-days agg.  & 17.8 (0.6)  & \textbf{15.9} (0.4)  \\ 
\hline
\end{tabular}}
\vspace{-10pt} 
\end{table}

\subsubsection{Role of policy data}
Table~\ref{tab:excluding_stringency_results} shows XGBoost's performance with the selected features (`Features'), and its performance when excluding the policy data (`w/o Stringency Index'). The findings show that the stringency index is a valuable feature for improving XGBoost's predictive performance. Notice that the third split is not considered in this evaluation, as Figure~\ref{fig:policies_3subplots} reveals that the index remains constant throughout the test period, thus being uninformative.

\begin{table}[t]
\centering
\caption{XGBoost's Perf. (MAPE) on Test Set using Selected Features Vs Excluding the Stringency Index}
\vspace{-5pt} 
\label{tab:excluding_stringency_results}
\renewcommand{\arraystretch}{1.2}
\setlength{\tabcolsep}{10pt}

\resizebox{\columnwidth}{!}{
\begin{tabular}{|c|c|c|c|}
\hline
\textbf{Split} & \textbf{MAPE (\%)} & 
\makecell{\textbf{Features}} & 
\makecell{\textbf{w/o Stringency Index}} \\ 
\hline
\multirow{2}{*}{1st} & 7-days avg  & \textbf{24.9} (2.1)  & 29.0 (1.9)  \\ 
& 7-days agg.  & \textbf{20.7} (0.9)  & 27.2 (1.4)  \\ 
\hline
\multirow{2}{*}{2nd} & 7-days avg  & 26.1 (0.7)  & \textbf{25.7} (0.5)  \\ 
& 7-days agg.  & \textbf{14.5} (0.4)  & 15.9 (0.1)  \\ 
\hline
\end{tabular}}
\vspace{-15pt} 
\end{table}

\subsubsection{Role of weather data}
Table~\ref{tab:excluding_weather_results} shows XGBoost's performance with the selected features (`Features'), and its performance when excluding the weather data (`w/o Weather'). The results indicate a notable decline in predictive performance when weather data is excluded. This reinforces the importance of weather data and confirming their necessity in our final feature set.

\begin{table}[t]
\centering
\caption{XGBoost's Perf. (MAPE) on Test Set using Selected Features Vs Excluding Weather Data}
\vspace{-8pt} 
\label{tab:excluding_weather_results}
\renewcommand{\arraystretch}{1.2}
\setlength{\tabcolsep}{16pt} 

\resizebox{\columnwidth}{!}{
\begin{tabular}{|c|c|c|c|}
\hline
\textbf{Split} & \textbf{MAPE (\%)} & 
\makecell{\textbf{Features}} & 
\makecell{\textbf{w/o Weather}} \\ 
\hline
\multirow{2}{*}{1st} & 7-days avg  & \textbf{24.9} (2.1)  & 33.7 (2.2)  \\ 
& 7-days agg.  & \textbf{20.7} (0.9)  & 24.5 (0.8)  \\ 
\hline
\multirow{2}{*}{2nd} & 7-days avg  & \textbf{26.1} (0.7)  & 26.5 (0.6)  \\ 
& 7-days agg.  & \textbf{14.5} (0.4)  & 15.0 (0.2)  \\ 
\hline
\multirow{2}{*}{3rd} & 7-days avg  & \textbf{25.0} (1.3)  & 25.9 (1.2)  \\ 
& 7-days agg.  & \textbf{17.8} (0.6)  & 18.8 (0.6)  \\ 
\hline
\end{tabular}}
\vspace{-13pt} 
\end{table}

\subsection{Important features for XGBoost and ARIMAX}
In this section, we compare the model performance when using only infections (raw), against using infections (raw + features), as shown in the first two categories of Table~\ref{tab:feature_categories}, and against the important features identified in previous experiments. These features include epidemiological data, the general stringency index, and selected weather variables. It excludes the vaccination-related data. We refer to these as `Important' features.

As shown in Table~\ref{tab:xgboost_infections_rawVsinfections_featVsimportant_feat}, which examines the performance of XGBoost by feature set selection, the first category, where only the window of positive cases is used, exhibits the worst results. Comparing the other categories, the first split shows substantial improvement when incorporating the important features, leading to a drastic reduction in the metrics. The second split produces similar results for the final two categories. Finally, the third split shows enhanced performance when important features are included.
Concluding, incorporating important features in XGBoost generally improves forecasting accuracy, highlighting their significance in enhancing the model.

As shown in Table~\ref{tab:arimax_infections_rawVsinfections_featVsimportant_feat}, for ARIMAX's performance, we observe that the last two categories perform similarly and consistently better than the raw infections.

\begin{table}[t]
\centering
\caption{XGBoost's Perf. (MAPE) by Feature Selection}
\vspace{-8pt} 
\label{tab:xgboost_infections_rawVsinfections_featVsimportant_feat}
\renewcommand{\arraystretch}{1.2}
\setlength{\tabcolsep}{7.5pt}

\resizebox{\columnwidth}{!}{
\begin{tabular}{|c|c|c|c|c|}
\hline
\textbf{Split} & \textbf{MAPE (\%)} &
\makecell{\textbf{Infections}\\\textbf{(Raw)}} &
\makecell{\textbf{Infections}\\\textbf{(Raw+Feat.)}} &
\makecell{\textbf{Important Feat.}\\\textbf{(excl.\ Vacc.)}} \\
\hline
\multirow{2}{*}{1st} & 7-days avg  & 28.1 (0.4) & 28.7 (1.2)  & \textbf{23.0} (1.2)  \\ 
& 7-days agg.  & 21.2 (0.3) & 25.6 (0.6)  & \textbf{17.7} (0.6)  \\ 
\hline
\multirow{2}{*}{2nd} & 7-days avg  & 27.9 (0.5) & \textbf{24.6} (1.1)  & 25.9 (0.7)  \\ 
& 7-days agg.  & 17.7 (0.2) & 15.8 (0.8)  & \textbf{14.6}(0.2)  \\ 
\hline
\multirow{2}{*}{3rd} & 7-days avg  & 27.9 (1.0) & 24.3 (1.0)  & \textbf{22.8} (0.9)  \\ 
& 7-days agg.  & 17.3 (0.3) & 17.6 (0.2)  & \textbf{15.9} (0.4)  \\ 
\hline
\end{tabular}}
\vspace{-10pt} 
\end{table}

\begin{table}[t]
\centering
\small
\caption{ARIMAX’s Perf.\ (MAPE) by Feature Selection}
\vspace{-8pt} 
\label{tab:arimax_infections_rawVsinfections_featVsimportant_feat}
\renewcommand{\arraystretch}{1.2}
\setlength{\tabcolsep}{7.5pt}

\resizebox{\columnwidth}{!}{
\begin{tabular}{|c|c|
                c|c|
                c|}
\hline
\textbf{Split} & \textbf{MAPE (\%)} &
\makecell{\textbf{Infections}\\\textbf{(Raw)}} &
\makecell{\textbf{Infections}\\\textbf{(Raw+Feat.)}} &
\makecell{\textbf{Important Feat.}\\\textbf{(excl.\ Vacc.)}} \\
\hline
\multirow{2}{*}{1st} 
  & 7-days avg & 30.4 & 18.9 & \textbf{17.7} \\
  & 7-days agg. & 26.7 & 16.0 & \textbf{13.6} \\
\hline
\multirow{2}{*}{2nd} 
  & 7-days avg & 29.7 & \textbf{22.7} & 23.3 \\
  & 7-days agg. & 23.9 & \textbf{17.7} & 18.1 \\
\hline
\multirow{2}{*}{3rd} 
  & 7-days avg & 25.2 & \textbf{19.6} & 21.9 \\
  & 7-days agg. & 16.0 & \textbf{13.1} & 15.1 \\
\hline
\end{tabular}
}
\vspace{-10pt}
\end{table}

\subsection{Role of the forecasting horizon for XGBoost and ARIMAX}

\begin{figure}[t]
    \centering
    \includegraphics[width=0.495\textwidth]{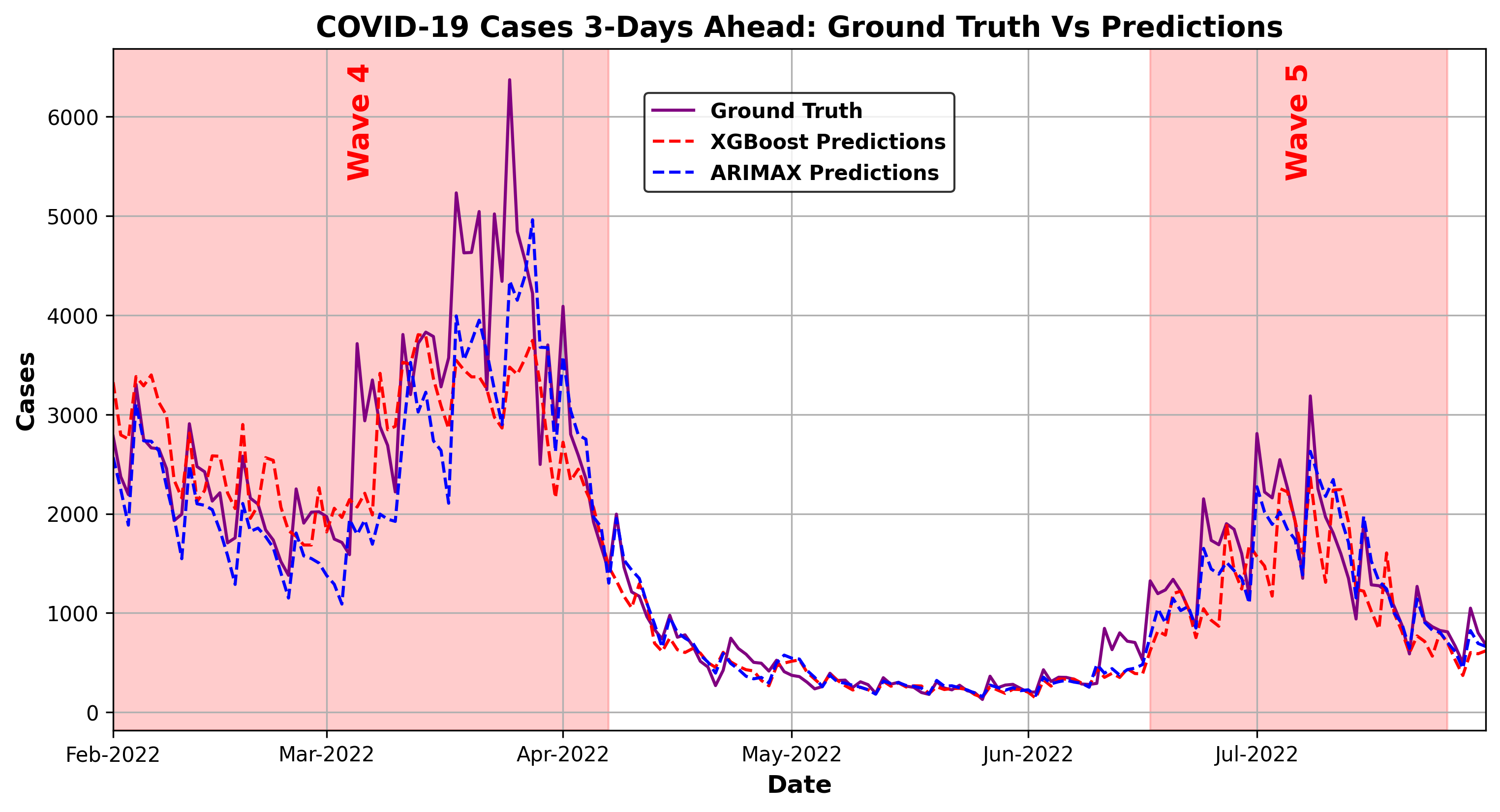}
    \vspace{-15pt}
    \caption{True vs. Predicted values for 3\textsuperscript{rd}-day-ahead forecasts.}
    \label{fig:forecasting_plot}
    \vspace{-15pt} 
\end{figure}

This section examines the effect of the forecasting horizon. Based on the findings of the previous section, XGBoost is evaluated using the important features, while ARIMAX is assessed using the infections (raw + features) category. 

Tables~\ref{tab:xgboost_arimax_comparison_split1_v3} and Table~\ref{tab:xgboost_arimax_comparison_split3} illustrate the results per model for the first and third split respectively. Importantly, as the forecasting horizon increases from 7-days to 14/21-days, XGBoost performs better during `wave' periods, while ARIMAX during `non-wave' periods. This is attributed to the fact that for periods without significant outbreaks (non-wave periods), the data may exhibit weak stationarity, while during outbreak `waves' the data dynamics undergo rapid and substantial changes which violate ARIMAX's stationarity assumption.

Figure~\ref{fig:forecasting_plot} presents the 3\textsuperscript{rd}-day-ahead predictions (i.e., predicting day \(d+3\) at day \(d\)), generated by XGBoost (averaged over 10 runs) and ARIMAX, respectively, compared against the actual values for our test set.

\begin{table*}[t]
\centering
\caption{Performance Comparison of XGBoost and ARIMAX across Different Periods (1st Split)}
\vspace{-8pt}
\label{tab:xgboost_arimax_comparison_split1_v3}
\renewcommand{\arraystretch}{1.2}
\setlength{\tabcolsep}{3pt}
\resizebox{\textwidth}{!}{
\begin{tabular}{|c|cc|cc||cc|cc||cc|cc|}
\hline
\multirow{2}{*}{\textbf{Target}}
 & \multicolumn{4}{c||}{\textbf{Overall Period}}
 & \multicolumn{4}{c||}{\textbf{Wave Period}}
 & \multicolumn{4}{c|}{\textbf{Non-wave Period}} \\
\cline{2-13}
 & \multicolumn{2}{c|}{\textbf{XGBoost}}
 & \multicolumn{2}{c||}{\textbf{ARIMAX}}
 & \multicolumn{2}{c|}{\textbf{XGBoost}}
 & \multicolumn{2}{c||}{\textbf{ARIMAX}}
 & \multicolumn{2}{c|}{\textbf{XGBoost}}
 & \multicolumn{2}{c|}{\textbf{ARIMAX}} \\
\cline{2-13}
 & \textbf{MAE} & \textbf{MAPE (\%)}
 & \textbf{MAE} & \textbf{MAPE (\%)}
 & \textbf{MAE} & \textbf{MAPE (\%)}
 & \textbf{MAE} & \textbf{MAPE (\%)}
 & \textbf{MAE} & \textbf{MAPE (\%)}
 & \textbf{MAE} & \textbf{MAPE (\%)} \\
\hline
\textbf{7-days avg}
 & 373 (24)  & 23.0 (1.2)
 & \textbf{317} & \textbf{18.9}
 & 551 (39)  & 22.9 (1.4)
 & \textbf{465} & \textbf{18.1}
 & 122 (9)   & 23.2 (2.0)
 & \textbf{107}        & \textbf{20.1} \\
\textbf{7-days agg}
 & 2072 (92) & 17.7 (0.6)
 & \textbf{1937} & \textbf{16.0}
 & 3114 (150) & 19.0 (0.8)
 & \textbf{2873} & \textbf{16.1}
 & \textbf{600} (36)  & \textbf{15.9} (0.7)
 & 615  & \textbf{15.9} \\ 
\hline
\textbf{14-days avg}
 & 467 (17)   & 32.3 (1.7)
 & \textbf{463} & \textbf{26.4}
 & \textbf{669} (26)   & 29.4 (1.2)
 & 689 & \textbf{27.7}
 & 181 (13)  & 36.6 (3.6)
 & \textbf{145}        & \textbf{24.5} \\
\textbf{14-days agg}
 & \textbf{5101} (139) & \textbf{22.5} (0.4)
 & 6020               & 23.7
 & \textbf{7405} (217) & \textbf{22.3} (0.2)
 & 9060               & 25.9
 & 1845 (63)         & 22.7 (0.8)
 & \textbf{1723}     & \textbf{20.7} \\
\hline
\textbf{21-days avg}
 & \textbf{566} (12)   & 47.8 (1.8)
 & 572                 & \textbf{31.6}
 & \textbf{763} (17)   & 39.8 (1.4)
 & 835                 & \textbf{33.7}
 & 288 (11)           & 59.0 (3.2)
 & \textbf{200}        & \textbf{28.7} \\
\textbf{21-days agg}
 & \textbf{9502} (52) & 32.8 (0.5)
 & 11305             & \textbf{28.9}
 & \textbf{12495} (60) & \textbf{25.9} (0.2)
 & 16726             & 31.5
 & 5271 (68)         & 42.5 (0.8)
 & \textbf{3644}     & \textbf{25.3} \\
\hline
\end{tabular}}
\vspace{-13pt}
\end{table*}

\begin{table*}[t]
\centering
\caption{Performance Comparison of XGBoost and ARIMAX across Different Periods (3rd Split)}
\vspace{-8pt}
\label{tab:xgboost_arimax_comparison_split3}
\renewcommand{\arraystretch}{1.2}
\setlength{\tabcolsep}{3pt} 
\resizebox{\textwidth}{!}{
\begin{tabular}{|c|cc|cc||cc|cc||cc|cc|}
\hline
\multirow{2}{*}{\textbf{Target}}
  & \multicolumn{4}{c||}{\textbf{Overall Period}}
  & \multicolumn{4}{c||}{\textbf{Wave Period}}
  & \multicolumn{4}{c|}{\textbf{Non‐wave Period}} \\
\cline{2-13}
  & \multicolumn{2}{c|}{\textbf{XGBoost}}
  & \multicolumn{2}{c||}{\textbf{ARIMAX}}
  & \multicolumn{2}{c|}{\textbf{XGBoost}}
  & \multicolumn{2}{c||}{\textbf{ARIMAX}}
  & \multicolumn{2}{c|}{\textbf{XGBoost}}
  & \multicolumn{2}{c|}{\textbf{ARIMAX}} \\
\cline{2-13}
 & \textbf{MAE} & \textbf{MAPE (\%)}
 & \textbf{MAE} & \textbf{MAPE (\%)}
 & \textbf{MAE} & \textbf{MAPE (\%)}
 & \textbf{MAE} & \textbf{MAPE (\%)}
 & \textbf{MAE} & \textbf{MAPE (\%)}
 & \textbf{MAE} & \textbf{MAPE (\%)} \\
\hline
\textbf{7‐days avg}
  & 127 (5)   & 22.8 (0.9)
  & \textbf{107} & \textbf{19.6}
  & \textbf{228} (25) & \textbf{17.0} (1.4)
  & 256       & 17.6
  & 97 (4)    & 24.0 (1.0)
  & \textbf{77}  & \textbf{20.0} \\

\textbf{7‐days agg}
  & 660 (21)  & 15.9 (0.4)
  & \textbf{557} & \textbf{13.1}
  & 1444 (101)& \textbf{12.5} (0.7)
  & \textbf{1391}   & 12.9
  & 500 (12)  & 16.6 (0.4)
  & \textbf{386}& \textbf{13.1} \\
\hline

\textbf{14‐days avg}
  & 156 (11)  & 33.2 (2.6)
  & \textbf{145} & \textbf{26.5}
  & \textbf{253} (32)& \textbf{20.2} (3.1)
  & 397       & 33.0
  & 136 (10)  & 35.8 (2.7)
  & \textbf{93}  & \textbf{25.2} \\

\textbf{14‐days agg}
  & \textbf{1564} (45) & 21.9 (0.6)
  & 1644               & \textbf{18.3}
  & \textbf{2207} (133)& \textbf{11.7} (0.8)
  & 4899               & 24.9
  & 1432 (40)         & 23.9 (0.6)
  & \textbf{977}      & \textbf{17.0} \\
\hline

\textbf{21‐days avg}
  & 185 (18)  & 43.5 (4.6)
  & \textbf{160} & \textbf{31.5}
  & \textbf{242} (38)& \textbf{27.3} (5.2)
  & 425       & 43.7
  & 173 (17)  & 46.8 (5.0)
  & \textbf{106} & \textbf{29.0} \\

\textbf{21‐days agg}
  & 2834 (64) & 28.9 (0.8)
  & \textbf{2779}& \textbf{21.5}
  & \textbf{2903} (185)& \textbf{13.8} (1.1)
  & 8225      & 32.7
  & 2820 (79)& 31.9 (0.9)
  & \textbf{1664}& \textbf{19.2} \\
\hline
\end{tabular}}
\vspace{-15pt}
\end{table*}

\vspace{-4pt}

\section{Conclusion}\label{sec:conclusion}

This study offers important empirical and methodological insights into COVID-19 forecasting by leveraging a multi-source dataset that incorporates epidemiological, policy, vaccination, and weather data, while focusing on a case study of Cyprus. Through the integration of diverse factors influencing infection trends, our work highlights the complex dynamics of the pandemic and the potential of data-driven approaches, along with external features, to support public health decision-making. While the findings are tailored to the Cypriot context, the methodology and lessons learned can inform efforts in other similar regions, contributing to broader pandemic preparedness and response frameworks.

Future work could explore the impact of features in predicting hospitalisations and ICU admissions, expanding the scope beyond infections to improve preparedness for future outbreaks. Additionally, recall that vaccination data in their aggregated form have been shown to have little predictive power. Future work will examine long-term vaccination effects and immunity development. Finally, hybrid models combining statistical and machine learning techniques may further enhance predictive accuracy.

\vspace{-4pt}

\def\IEEEbibitemsep{0pt plus .5pt}
\def\IEEEbibitemindent{0pt}
\bibliographystyle{IEEEtran}

\bibliography{investigating_forecasting_models_for_pandemic_infections_using_heterogeneous_data_sources_a_two_year_study_with_covid19}

\begin{thebibliography}{10}
\providecommand{\url}[1]{#1}
\csname url@samestyle\endcsname
\providecommand{\newblock}{\relax}
\providecommand{\bibinfo}[2]{#2}
\providecommand{\BIBentrySTDinterwordspacing}{\spaceskip=0pt\relax}
\providecommand{\BIBentryALTinterwordstretchfactor}{4}
\providecommand{\BIBentryALTinterwordspacing}{\spaceskip=\fontdimen2\font plus
\BIBentryALTinterwordstretchfactor\fontdimen3\font minus \fontdimen4\font\relax}
\providecommand{\BIBforeignlanguage}[2]{{%
\expandafter\ifx\csname l@#1\endcsname\relax
\typeout{** WARNING: IEEEtran.bst: No hyphenation pattern has been}%
\typeout{** loaded for the language `#1'. Using the pattern for}%
\typeout{** the default language instead.}%
\else
\language=\csname l@#1\endcsname
\fi
#2}}
\providecommand{\BIBdecl}{\relax}
\BIBdecl

\bibitem{who2019}
\BIBentryALTinterwordspacing
WHO. Emergencies: International health regulations and emergency committees. Accessed 13 Mar. 2025. [Online]. Available: \url{https://www.who.int/news-room/questions-and-answers/item/emergencies-international-health-regulations-and-emergency-committees}
\BIBentrySTDinterwordspacing

\bibitem{jones2008global}
K.~E. Jones, N.~G. Patel, M.~A. Levy, A.~Storeygard, D.~Balk, J.~L. Gittleman, and P.~Daszak, ``Global trends in emerging infectious diseases,'' \emph{Nature}, vol. 451, no. 7181, pp. 990--993, 2008.

\bibitem{coronaviridae2020species}
C.~S.~G. of~the International Committee on Taxonomy~of Viruses, ``The species severe acute respiratory syndrome-related coronavirus: classifying 2019-ncov and naming it sars-cov-2,'' \emph{Nature microbiology}, vol.~5, no.~4, pp. 536--544, 2020.

\bibitem{who2023}
\BIBentryALTinterwordspacing
WHO. Statement on the fifteenth meeting of the ihr (2005) emergency committee on the covid-19 pandemic. Accessed 13 Mar. 2025. [Online]. Available: \url{https://www.who.int/news/item/05-05-2023-statement-on-the-fifteenth-meeting-of-the-international-health-regulations-(2005)-emergency-committee-regarding-the-coronavirus-disease-(covid-19)-pandemic}
\BIBentrySTDinterwordspacing

\bibitem{ayoobi2021time}
N.~Ayoobi, D.~Sharifrazi, R.~Alizadehsani, A.~Shoeibi, J.~M. Gorriz, H.~Moosaei, A.~Khosravi, S.~Nahavandi, A.~G. Chofreh, F.~A. Goni \emph{et~al.}, ``Time series forecasting of new cases and new deaths rate for covid-19 using deep learning methods,'' \emph{Results in physics}, vol.~27, p. 104495, 2021.

\bibitem{nesteruk2020simulations_sir}
I.~Nesteruk, ``Simulations and predictions of covid-19 pandemic with the use of sir model,'' 2020.

\bibitem{pandey2020seir}
G.~Pandey, P.~Chaudhary, R.~Gupta, and S.~Pal, ``Seir and regression model based covid-19 outbreak predictions in india,'' \emph{arXiv preprint arXiv:2004.00958}, 2020.

\bibitem{karapitta2024time}
M.~Karapitta, A.~Kasis, C.~Stylianides, K.~Malialis, and P.~Kolios, ``Time-varying compartmental models with neural networks for pandemic infection forecasting,'' in \emph{2024 46th Annual International Conference of the IEEE Engineering in Medicine and Biology Society (EMBC)}.\hskip 1em plus 0.5em minus 0.4em\relax IEEE, 2024, pp. 1--5.

\bibitem{karapitta2024pandemic}
------, ``Pandemic infection forecasting through compartmental model and learning-based approaches,'' \emph{arXiv preprint arXiv:2401.06629}, 2024.

\bibitem{alzahrani2020forecasting_arma_arima}
S.~I. Alzahrani, I.~A. Aljamaan, and E.~A. Al-Fakih, ``Forecasting the spread of the covid-19 pandemic in saudi arabia using arima prediction model under current public health interventions,'' \emph{Journal of infection and public health}, vol.~13, no.~7, pp. 914--919, 2020.

\bibitem{benvenuto2020application_arima}
D.~Benvenuto, M.~Giovanetti, L.~Vassallo, S.~Angeletti, and M.~Ciccozzi, ``Application of the arima model on the covid-2019 epidemic dataset,'' \emph{Data in brief}, vol.~29, p. 105340, 2020.

\bibitem{ryan2024_non_stationarity}
O.~Ryan, J.~M. Haslbeck, and L.~J. Waldorp, ``Non-stationarity in time-series analysis: Modeling stochastic and deterministic trends,'' \emph{Multivariate Behavioral Research}, pp. 1--33, 2024.

\bibitem{ogundokun2020predictive_linear_regression}
R.~O. Ogundokun, A.~F. Lukman, G.~B. Kibria, J.~B. Awotunde, and B.~B. Aladeitan, ``Predictive modelling of covid-19 confirmed cases in nigeria,'' \emph{Infectious Disease Modelling}, vol.~5, pp. 543--548, 2020.

\bibitem{fang2022application_xgboost}
Z.-g. Fang, S.-q. Yang, C.-x. Lv, S.-y. An, and W.~Wu, ``Application of a data-driven xgboost model for the prediction of covid-19 in the usa: a time-series study,'' \emph{BMJ open}, vol.~12, no.~7, p. e056685, 2022.

\bibitem{chimmula2020time_lstm}
V.~K.~R. Chimmula and L.~Zhang, ``Time series forecasting of covid-19 transmission in canada using lstm networks,'' \emph{Chaos, solitons \& fractals}, vol. 135, p. 109864, 2020.

\bibitem{shahid2020predictions_comparative1}
F.~Shahid, A.~Zameer, and M.~Muneeb, ``Predictions for covid-19 with deep learning models of lstm, gru and bi-lstm,'' \emph{Chaos, Solitons \& Fractals}, vol. 140, p. 110212, 2020.

\bibitem{zeroual2020deep_comparative2}
A.~Zeroual, F.~Harrou, A.~Dairi, and Y.~Sun, ``Deep learning methods for forecasting covid-19 time-series data: A comparative study,'' \emph{Chaos, solitons \& fractals}, vol. 140, p. 110121, 2020.

\bibitem{malialis2020online}
K.~Malialis, C.~G. Panayiotou, and M.~M. Polycarpou, ``Online learning with adaptive rebalancing in nonstationary environments,'' \emph{IEEE transactions on neural networks and learning systems}, vol.~32, no.~10, pp. 4445--4459, 2020.

\bibitem{stylianides2023study}
C.~Stylianides, K.~Malialis, and P.~Kolios, ``A study of data-driven methods for adaptive forecasting of covid-19 cases,'' in \emph{International Conference on Artificial Neural Networks}.\hskip 1em plus 0.5em minus 0.4em\relax Springer, 2023, pp. 62--74.

\bibitem{tetteroo2022automated_onlineextra1}
J.~Tetteroo, M.~Baratchi, and H.~H. Hoos, ``Automated machine learning for covid-19 forecasting,'' \emph{IEEE Access}, vol.~10, pp. 94\,718--94\,737, 2022.

\bibitem{vaccinations}
\BIBentryALTinterwordspacing
N.~O. D. P. o.~C. Ministry~of Health, Government of~Cyprus. Weekly vaccination statistics against covid-19 (by target group). Accessed 16 Oct. 2024. [Online]. Available: \url{https://data.gov.cy/el/dataset/ebdomadiaia-statistika-emboliasmon-kata-tis-nosoy-covid-19-ana-omada-stoho}
\BIBentrySTDinterwordspacing

\bibitem{policies}
\BIBentryALTinterwordspacing
T.~Hale, N.~Angrist, R.~Goldszmidt, B.~Kira, A.~Petherick, T.~Phillips, S.~Webster, E.~Cameron-Blake, L.~Hallas, S.~Majumdar \emph{et~al.}, ``A global panel database of pandemic policies (oxford covid-19 government response tracker),'' \emph{Nature human behaviour}, vol.~5, no.~4, pp. 529--538, 2021, accessed 24 July 2024. [Online]. Available: \url{https://doi.org/10.1038/s41562-021-01079-8}
\BIBentrySTDinterwordspacing

\bibitem{weather_data}
\BIBentryALTinterwordspacing
V.~Crossing. Weekly vaccination statistics against covid-19 (by target group). Accessed 29 Aug. 2024. [Online]. Available: \url{https://www.visualcrossing.com/weather-query-builder/}
\BIBentrySTDinterwordspacing

\bibitem{chen2016xgboost}
T.~Chen and C.~Guestrin, ``Xgboost: A scalable tree boosting system,'' in \emph{Proceedings of the 22nd acm sigkdd international conference on knowledge discovery and data mining}, 2016, pp. 785--794.

\bibitem{box2013box}
G.~Box, ``Box and jenkins: time series analysis, forecasting and control,'' in \emph{A Very British Affair: Six Britons and the Development of Time Series Analysis During the 20th Century}.\hskip 1em plus 0.5em minus 0.4em\relax Springer, 2013, pp. 161--215.

\bibitem{guyon2002gene_rfe}
I.~Guyon, J.~Weston, S.~Barnhill, and V.~Vapnik, ``Gene selection for cancer classification using support vector machines,'' \emph{Machine learning}, vol.~46, pp. 389--422, 2002.

\bibitem{cdc2024}
\BIBentryALTinterwordspacing
C.~for Disease~Control and P.~(CDC). (2024) Covid-19 vaccine basics. Accessed 23 Mar. 2025. [Online]. Available: \url{https://www.cdc.gov/covid/vaccines/how-they-work.html}
\BIBentrySTDinterwordspacing

\bibitem{zeng2022effectiveness}
B.~Zeng, L.~Gao, Q.~Zhou, K.~Yu, and F.~Sun, ``Effectiveness of covid-19 vaccines against sars-cov-2 variants of concern: a systematic review and meta-analysis,'' \emph{BMC medicine}, vol.~20, no.~1, p. 200, 2022.

\bibitem{who_vaccinations_2021}
\BIBentryALTinterwordspacing
WHO. (2021) Vaccine efficacy, effectiveness and protection. Accessed 23 Mar. 2025. [Online]. Available: \url{https://www.who.int/news-room/feature-stories/detail/vaccine-efficacy-effectiveness-and-protection}
\BIBentrySTDinterwordspacing

\end{thebibliography}

\end{document}